\documentclass[conference]{IEEEtran}
\IEEEoverridecommandlockouts
\usepackage{cite}
\usepackage{amsmath,amssymb,amsfonts}
\usepackage{dblfloatfix} 
\usepackage{graphicx}
\usepackage{textcomp}
\usepackage{xcolor}
\usepackage[perpage]{footmisc}
\def\BibTeX{{\rm B\kern-.05em{\sc i\kern-.025em b}\kern-.08em
    T\kern-.1667em\lower.7ex\hbox{E}\kern-.125emX}}
\begin{document}

\title{Music Artist Classification with Convolutional Recurrent Neural Networks}

\author{\IEEEauthorblockN{Zain Nasrullah}
\IEEEauthorblockA{\textit{Department of Computer Science} \\
\textit{University of Toronto}\\
Toronto, Canada \\
znasrullah@cs.toronto.edu}
\and
\IEEEauthorblockN{Yue Zhao}
\IEEEauthorblockA{\textit{Department of Computer Science} \\
\textit{University of Toronto}\\
Toronto, Canada \\
yuezhao@cs.toronto.edu}
}

\maketitle

\begin{abstract}
Previous attempts at music artist classification use frame level audio features which summarize frequency content within short intervals of time. Comparatively, more recent music information retrieval tasks take advantage of temporal structure in audio spectrograms using deep convolutional and recurrent models. This paper revisits artist classification with this new framework and empirically explores the impacts of incorporating temporal structure in the feature representation. To this end, an established classification architecture, a Convolutional Recurrent Neural Network (CRNN), is applied to the \textit{artist20} music artist identification dataset under a comprehensive set of conditions. These include audio clip length, which is a novel contribution in this work, and previously identified considerations such as dataset split and feature level. Our results improve upon baseline works, verify the influence of the producer effect on classification performance and demonstrate the trade-offs between audio length and training set size. The best performing model achieves an average F1 score of 0.937 across three independent trials which is a substantial improvement over the corresponding baseline under similar conditions. Additionally, to showcase the effectiveness of the CRNN's feature extraction capabilities, we visualize audio samples at the model's bottleneck layer demonstrating that learned representations segment into clusters belonging to their respective artists.

\end{abstract}


\section{Introduction}

Music information retrieval (MIR) encompasses most audio analysis tasks such as genre classification, song identification, chord recognition, sound event detection, mood detection and feature extraction \cite{choi2016explaining, mcfee2015librosa}. Search algorithms in particular, popularized by Shazam Entertainment, are able to efficiently identify a song and the performing artist given an audio sample \cite{wang2003industrial}. These methods, however, require a database to match patterns against and thus cannot generalize beyond known songs. In contrast, humans are able to identify the artist performing a new song if familiar with the artist in question. This ability to process and identify with music has roots in neuroscience \cite{panksepp2009neuroscience} and thus an algorithmic representation would work best if it included an analogous learning model.
    
While previous attempts at artist classification with machine learning exist, they use low-dimensional feature representations to summarize the audio signal \cite{whitman2001artist, labrosa20}. Historically, this was a necessary compromise because traditional learning models suffered from the curse of dimensionality \cite{verleysen2005curse} and lacked the necessary compute resources to learn from large amounts of high-dimensional data.  Artist classification is also particularly challenging because a limited number of training examples exist for each artist; furthermore, these examples vary stylistically from one song to another and as an artist's style changes over time. Although the task is still manageable for humans given a small number of artists, it becomes much more difficult as the number of artists begins to scale---one would need to be intimately familiar with the discography of dozens of artists. Therefore, successfully modelling artist classification could yield benefits for automation in the music industry. For example, such a model could be used to detect copyright violations or identify songs that sound stylistically similar to another artist as part of a recommendation system. 

Since neural networks are inspired by how the mind works \cite{hinton1992neural} and Convolutional Neural Networks (CNN) have also been shown to model the way humans visually process information \cite{lecun2015deep}, deep learning models would be ideal for a learning-based approach to artist classification. A key distinction between MIR and traditional tasks in deep learning is that a good representation of auditory data is difficult to identify. As mentioned, past research favored vector summaries of frequency content in an audio window, specifically Mel-frequency cepstral coefficients (MFCCs), because they summarize both key and timbral information in an audio track \cite{li2011genre}. The problem with this and other established approaches \cite{tzanetakis2002musical} is that, in summarizing a short-duration of audio into a vector, temporal structure is lost. Recent works by Choi et al. \cite{choi2016automatic} and Cakir et al. \cite{ccakir2017convolutional} demonstrate that addressing this issue by using audio spectrograms, which contain frequency content over time, as a feature representation achieves promising results in genre tagging and sound event detection. Representing audio as a spectrogram allows convolutional layers to learn global structure and recurrent layers to learn temporal structure.

To that end, this study adapts the Convolutional Recurrent Neural Network (CRNN) architecture proposed for audio tasks in prior work \cite{choi2017convolutional} to establish a baseline for artist classification with deep learning. Overall, we aim to validate whether the deep learning model is comparable to historical methods of artist classification despite the limitation on the number of songs per artist. As part of this process, we investigate the effect of audio length on classification performance both when the dataset is split by album and by song. This consideration is recognized \cite{whitman2001artist} as important in characterizing performance. We also validate whether aggregating frame level predictions up to the parent song, as in a prior work \cite{mandel2005song}, yields noise reduction and improves performance. To foster reproducibility, all code, figures and results are openly shared\footnote{https://github.com/ZainNasrullah/music-artist-classification-crnn}.

As a whole, this work makes the following contributions: 
\begin{enumerate}
    \item To our knowledge, this is the first comprehensive study of deep learning applied to music artist classification. It explores six audio clip lengths, an album versus song data split and frame level versus song level evaluation producing results under twenty different conditions.  
    \item Our results outperform the most comparable baselines (in terms of split type and feature level) under at least one audio clip length.
    \item We visualize audio samples at the bottleneck layer of the CRNN to showcase that learned representations form clusters belonging to their respective artist. 
\end{enumerate}

\section{Related Works}
    \begin{figure*} [ht]
    \centering
    \includegraphics[width=1\textwidth]{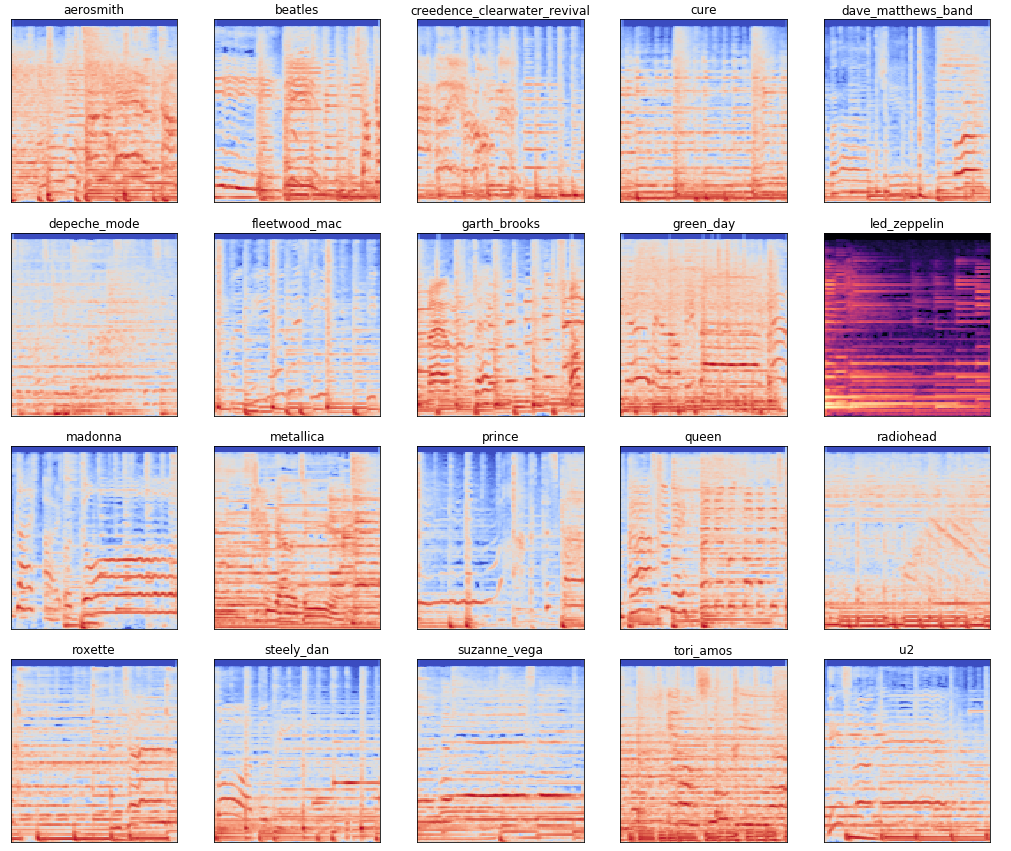}
    \caption{\label{fig:artists}Spectrograms for all artists in the \textbf{\textit{artist20}} dataset (songs randomly chosen; duration three seconds).}
    \label{artists}
    \end{figure*}
    
    \subsection{Machine Learning in Music Related Tasks}
    Current research with auditory data focuses on areas such as sound event detection, speech transcription or speech generation \cite{van2016wavenet, deng2014deep, lecun2015deep} and, leveraging deep learning, massive performance gains have been achieved in these areas. With the explosive shift towards cloud-based platforms in the music industry, the relevancy of machine learning for automating, categorizing and personalizing services has also grown. Particularly, substantial attention is being given to research in music recommendation \cite{volkovs2018two} and genre classification \cite{choi2016explaining}. The benefit of approaching these problems with deep learning is that neural networks excel as feature extractors \cite{lee2009unsupervised} and thus learn representations which are easier to classify compared to prior approaches using summarized frame level features.
    
    While these results showcase the effectiveness of deep learning in MIR, almost no attention is given to the similar task of artist classification in spite of its potential in creating effective audio representations. Unlike genre classification, where the ground truth may be considered subjective \cite{lippens2004comparison}, artist classification possesses an objective ground truth; therefore, its learned representations may be more meaningful for augmenting other tasks such as recommendation systems. Learning the style of an artist is also a good method for identifying copyright violations by comparing the similarity of a new audio sample to previously learned representations.
    
    \subsection{Artist Classification Baselines} \label{baseline}
    Advances in artist classification with machine learning are limited. Furthermore, while exploratory attempts exist \cite{nibbeling}, a comprehensive study with deep learning is still absent. Prior academic works are almost a decade old and employ traditional algorithms which do not work well with high-dimensional and sequential data. 
    
    Whitman et al. \cite{whitman2001artist}, for example, use a one second MFCC feature representation and a Support Vector Machine (SVM) classification model to achieve a best test accuracy of 50\%. While the dataset used in their study has not been released, the authors state that it contains a mix of multiple genres across 240 songs. A limitation mentioned in this foundational work is that it does not address the producer effect which refers to the model learning patterns related to the production of an album in addition to the artist's musical style. This inflates evaluation metrics if the dataset is split by song since the test set may contain songs from the same album as the training set. Splitting the dataset by album, where an over-reliance on production level details is exposed during evaluation, results in lower scores. Given that the type of split depends on the intended use case of the model, this work explores both splits to evaluate performance under a variety of conditions.  
    
    Addressing this issue alongside the release of the \textbf{\textit{artist20}} dataset, Labrosa uses a full-covariance Gaussian classifier to establish an artist classification baseline \cite{labrosa20} with an album split. Using randomly sampled MFCC vectors from each artist, the model achieves 56\% accuracy. By including additional hand-crafted features, the final model achieves a best accuracy of 59\%. The authors acknowledge that better performance may have been achieved by ensembling predictions at the song level but choose not to explore that avenue. 
    
    Mandel and Ellis \cite{mandel2005song} study the impact of frame level versus song level evaluation using Gaussian Mixture Models (GMM) and SVMs. On the \textbf{\textit{uspop2002}} dataset, at the frame level, the GMM achieves comparable performance (54\% accuracy) to Labrosa's baseline \cite{labrosa20}. At the song level, the SVM approach obtains best accuracies of 68.7\% and 83.9\% with an album and song dataset split respectively. These results validate that song level evaluation outperforms frame level and that the producer effect has a strong impact on performance. 
    
    
    All of the works discussed in this subsection are summarized in Table \ref{baselines_table} and used as baselines in this study.

    \subsection{Audio Representations and Spectrograms}
    Traditional MIR tasks relied heavily on MFCCs which extract frequency content in a short-window frame of audio; however, recent works are shifting towards spectrograms which take advantage of temporal structure and have shown better performance in classification tasks \cite{choi2016automatic}. A spectrogram is a representation of frequency content over time found by taking the squared magnitude of the short-time Fourier Transform (STFT) of a signal \cite{mcfee2015librosa}. The mathematical form of the discrete STFT is shown in Eq. \eqref{stftdiscrete}, where $x[n]$ and $w[n]$ describe the input signal and window function \cite{pielemeier1996time}. 
    
    \begin{equation}\label{stftdiscrete}
    STFT\{x(n)\}(m, \omega) = \sum_{n=-\infty}^{\infty}{x[n]w[n-m]e^{-j\omega n}}
    \end{equation}
     
    Figure \ref{artists} illustrates how a spectrogram captures both frequency content and temporal variation for three second audio samples in the \textbf{\textit{artist20}} dataset. Although spectrograms can be difficult to interpret, simple audio features such as the presence of sound and its frequency range are readily identifiable. With familiarity, it is possible to extract more information from this representation such as identifying which instruments are being played \cite{choi2016explaining}. We hypothesize that the patterns across frequency and time also contain stylistic tendencies associated with an artist and thus deep learning architectures, such as Convolutional Neural Networks which excel at pattern recognition in two-dimensional data, should be able to learn them.
    
    
    \subsection{Convolutional Recurrent Neural Networks}
    
     Since their early successes on ImageNet \cite{krizhevsky2012imagenet}, CNNs have become a standard in deep learning when working with visual data. In a prior work, Choi et al. \cite{choi2016explaining} discuss how convolution can be used in MIR tasks; notably, they demonstrate that the layers in a CNN act as feature extractors. Generally, low-level layers describe sound onsets or the presence of individual instruments while high-level layers describe abstract patterns. This is consistent with established work \cite{lecun2015deep} which suggests that deep convolutional layers are a composition of lower-level features. Recurrent Neural Networks (RNN) have also had success with audio tasks such as speech recognition \cite{graves2013speech} because sound is sequential along the time axis. In a follow-up study, Choi et al. \cite{choi2017convolutional} show that including a recurrent unit to summarize temporal structure following 2D convolution, dubbed Convolutional Recurrent Neural Network, achieves the best performance in genre classification among four well-known audio classification architectures. In this work, we adapt the CRNN model to establish a deep learning baseline for artist classification.

\section{Methodology}

    \subsection{Dataset} \label{dataset}
    The music artist identification dataset \textbf{\textit{artist20}}\footnote{Available at https://labrosa.ee.columbia.edu/projects/artistid/}, created by Labrosa \cite{labrosa20}, is used to evaluate classification performance. It contains six albums from twenty artists covering a range of musical styles. A summary is provided in Table \ref{Artist20Specs} and the distinctiveness between artists is highlighted in Figure \ref{artists}. 
   
       \begin{table}
        \caption{Artist20 Dataset Specifications}
        \centering
        \begin{tabular}{|c|c|}
        \hline
        \textbf{Property}                & Value   \\
        \hline
        \hline
        \textbf{Total Number of Tracks}  & 1,413   \\
        \textbf{Total Number of Artists} & 20      \\
        \textbf{Albums Per Artist}       & 6       \\
        \textbf{Bitrate}                 & 32 kbps \\
        \textbf{Sample Rate}             & 16 kHz  \\
        \textbf{Channels}                & Mono    \\
        \hline
        \end{tabular}
        \label{Artist20Specs}
    \end{table}
    
    The train-test split in artist classification is an important consideration. In addition to ensuring frames from test songs are not used in training, one must also account for the producer effect identified in prior work by Whitman et al. \cite{whitman2001artist}. This refers to inflated classification performance in datasets split by song because of how salient production details can be in comparison to musical style. To combat this, the standard approach is to split the dataset by album such that the test set is composed solely of songs from albums not used in training. However, any model trained under this paradigm would not be robust to changes in musical styles across albums. Furthermore, production level details associated with an album could also be considered part of an artist's unique style. For these reasons, we explore both song and album train-test splits and compare our results to a prior work by Mandel and Ellis \cite{mandel2005song} which also takes a similar approach. It is worth noting that their work uses an eighteen artist subset of the \textbf{\textit{uspop2002}} dataset which evolved into \textbf{\textit{artist20}}. Therefore, although their evaluation contains fewer artists, it is still a reasonable baseline because of the substantial overlap in the dataset. 
    
    In this work, spectrograms are created for the entire length of each song to form an initial dataset. For the song split, this dataset undergoes a 90/10 stratified-by-artist split to create train and test sets respectively. The train set is then split using the same stratified 90/10 split to create train and validation (used for early stopping) subsets. The stratification ensures that each set contains an equivalent number of songs from each artist. For the album split, two albums from each artist are randomly removed from the initial dataset---one is added to the test set and the other to validation. The remaining four albums from each artist are joined to form the training set. 

    \subsection{Audio Processing} \label{audio-processing}
    
    A short-time Fourier transform is applied to the raw audio signal for every song to create spectrograms. Once created, the frequency scale ($f$ hertz) is transformed into the Mel scale ($m$ mels) using Eq. \eqref{mel-scale} and then scaled ($d$ decibels) using Eq. \eqref{log-amplitude}.
    \begin{equation} \label{mel-scale}
        m = 2595\log_{10}(1+f/700)
    \end{equation}
    
    \begin{equation} \label{log-amplitude}
        d = 10\log_{10}(m/r)
    \end{equation}
    
    These operations are considered standard practices for audio processing and have been shown in prior work \cite{choi2016automatic} to improve performance in classification tasks. The parameters used throughout this process are summarized in Table \ref{stft} and are based on best practices specified in documentation \cite{mcfee2015librosa} and earlier research \cite{van2013deep}. The only exception to this is the sampling rate which is set at the specified value for audio tracks in the \textbf{\textit{artist20}} dataset.
    
    \begin{table}
    \caption{FFT Specifications}
    \centering
    \begin{tabular}{|c|c|}
    \hline
    \textbf{Property}                        & Value  \\
    \hline
    \hline
    \textbf{Sampling Rate}                   & 16 kHz \\
    \textbf{Number of Mel Bins}              & 128    \\
    \textbf{FFT Window Size}                 & 2048   \\
    \textbf{Hop Length}                      & 512    \\
    \textbf{Reference Power for Log-Scaling ($r$)} & 1.0    \\
    \hline
    \end{tabular}
    \label{stft}
    \end{table}
    
    \begin{figure*} [ht]
    \centering
    \includegraphics[width=1\textwidth]{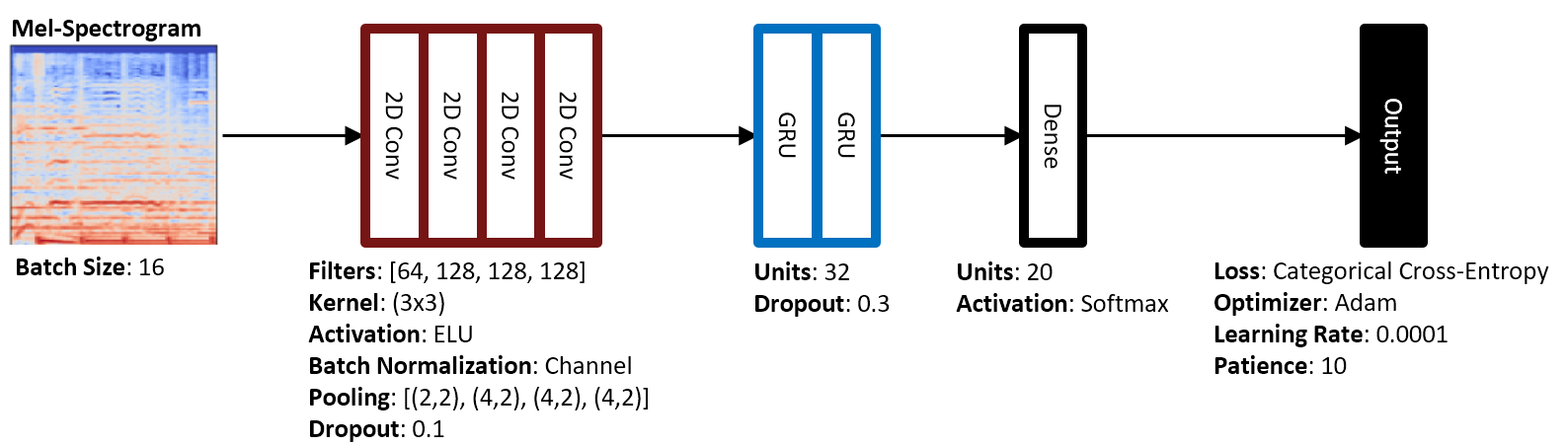}
    \caption{\label{fig:CRNN}CRNN Architecture for Artist Classification} \label{CRNN}
    \end{figure*}   
    
    The spectrograms of each song are split into train, test and validation sets as described in Section \ref{dataset}. Once split, each song is sliced into audio clips of length $t$, which is varied in the study. Furthermore, rather than use a single clip per song, the entirety of each song is used during training and evaluation. This is in contrast to prior MIR works using CNNs \cite{choi2017convolutional} where only a highlight audio clip is used for each song. The benefit of taking our approach is that it yields a greater number of training samples and allows for experimentation with song level predictions. Furthermore, there is a trade-off between the training set size and the length of each audio clip. Longer clips contain more temporal structure within each training sample while shorter clips can be shuffled and interpreted as a larger set of independent training samples. This trade-off is discussed further in Section \ref{baseline_performance}.
    
    \subsection{Frame Level versus Song Level Evaluation} \label{evaluation_method}
    
    Prior works discussed in Section \ref{baseline} also vary with respect to how the model is trained and correspondingly evaluated. Classification can be performed at the individual frame level, where each frame is treated as an independent sample, or at the song level where the goal is to classify the artist corresponding to a particular song using multiple samples. The latter can be interpreted as a form of ensembling where aggregating frame level predictions and voting up to the song level can yield variance reduction; this has been an effective approach in various interdisciplinary machine learning studies \cite{zhao2018xgbod,zhao2018employee,zhao2019lscp}. 
    
    To compare with all baselines, this work investigates both evaluation strategies. Specifically, at the frame level, each $t$-length audio spectrogram is treated as an independent sample and performance is measured by taking the F1 score across all samples in the test set. To evaluate at the song level, the most frequent frame level artist prediction belonging to each song is selected as the final prediction. The F1 score is then reported by song to quantify performance. 
    
    The F1 score is used, instead of accuracy, because all audio slices within each song are used during training and evaluation. Therefore, minor class imbalance results from variance in song length: artists who frequently make longer or shorter songs compared to the average song length will have an imbalanced number of training examples. This is in contrast to our baselines which random sample a fixed number of MFCC vectors from each song to form train and test sets. Weighting the F1 score by the number of supporting samples in each class helps mitigate the impact of class imbalance during evaluation while allowing us to make use of all training data.
    
    \subsection{Model Architecture and Design}
    
    Since our experiments in artist classification aim to replace the traditional MFCC-based approaches with high-dimensional spectrograms, we adapt the CRNN architecture from previous work \cite{choi2017convolutional} in genre classification. It is expected that this architecture would also work well for artist classification because understanding musical style involves characterizing how frequency content changes over time. Given that this information is contained within a spectrogram, the ideal network architecture must be able to summarize patterns in frequency (where convolutional layers excel) and then also consider any resulting temporal sequences in these patterns (where recurrent layers excel). The CRNN contains both of these components. Alternate models and hyper-parameters were tested, but did not show substantial performance gain for the computational cost of expanding the network and are thus excluded from the results presented in this paper. The final model is described in Figure \ref{CRNN} and rationale for design choices follow. The architecture can broadly be divided up into three stages: convolutional, recurrent and fully-connected. 
    
    The number of layers, number of filters and the kernel size in the convolutional component are adapted directly from prior work \cite{choi2017convolutional}. The Exponential Linear Unit (ELU) activation function is used as a smooth alternative of the Rectified Linear Unit (RELU) because it has shown better generalization performance\cite{clevert2015fast}. Batch normalization (normalizing across channels) and drop-out (regularization) are also included to improve generalization as per earlier research in image classification tasks \cite{sigtia2014improved,ioffe2015batch}. Pooling and stride are selected to fully characterize the 128-bin frequency axis and summarize the sequence length entering the recurrent layer. 
    
    The recurrent component replaces the need for explicit temporal pooling by acting as a form of temporal summarization. Gated Recurrent Units (GRUs) are used for this purpose instead of Long Short Term Memory (LSTM) cells since they require fewer parameters and have similar performance \cite{chung2014empirical}. The final fully-connected layer assigns probabilities to each class with a softmax activation.

    \subsection{Training Considerations}
    
    Artist identification can be treated as a multi-class classification problem; correspondingly, categorical cross-entropy is selected as the loss function in our experiments. Adam is chosen as the optimizer because it has shown state of the art performance in convolution-based classification tasks with limited hyper-parameter tuning \cite{kingma2014adam}. However, the default learning rate (0.001) is reduced an order of magnitude to improve training stability. Early stopping is also used, with a patience of 10, to mitigate overfitting.

\section{Results and Discussion}
   
   This section discusses the results of all experiments and compares them to baselines (summarized in Table \ref{baselines_table}) established in prior works. The CRNN model used in this study is trained with audio clips of length \{1s, 3s, 5s, 10s, 20s, 30s\} under various conditions such as split-type \{song, album\} and feature level \{frame, song\}. The average and best test F1 scores across three independent runs are summarized in Tables \ref{results_frame} and \ref{results_song} for frame and song level results respectively. The F1 score is reported since the data is not balanced, given that artists with longer songs will have more training samples available, and is thus a better measure of performance than accuracy (see Section \ref{evaluation_method} for more details). For comparative purposes, while it is acknowledged that accuracy and the F1 score are not equivalent, the metrics are regarded as representative indicators of performance.   
    
    \begin{table}
    \caption{Test Accuracy Scores for MFCC Baselines (Best Reported)} 
    \centering
    \begin{tabular}{|l|ccccc|}
    \hline
    \textbf{Work} & \textbf{Level} & \textbf{Split} & \textbf{Artists} & \textbf{Method} & \textbf{Accuracy} \\
    \hline
    \hline
    Whitman\cite{whitman2001artist} & Frame & Song  & 21 & SVM   & 0.500  \\
    Labrosa\cite{labrosa20}         & Frame & Album & 20 & GMM      & 0.590  \\
    Mandel \cite{mandel2005song}     & Frame & Album & 18 & GMM      & 0.541  \\
    Mandel \cite{mandel2005song}     & Song  & Album & 18 & SVM  & 0.687  \\
    Mandel \cite{mandel2005song}     & Song  & Song  & 18 & SVM  & 0.839  \\
    \hline
    \end{tabular}
    \label{baselines_table}
    \end{table}
 
    \subsection{Frame Level Evaluation} \label{baseline_performance}

    The best baseline frame level performance is achieved by Labrosa's \cite{labrosa20} Gaussian Mixture Model (accuracy of 0.590). Comparatively, for an album split, our results (see Table  \ref{results_frame}) at one second of audio are initially disappointing but improve as temporal structure is added to the feature representation. At three seconds, performance exceeds the SVM by Whitman et al. and begins to approach Labrosa's result. Finally, at thirty seconds, our average and best F1 scores of 0.603 and 0.612 respectively showcase the benefit of the spectrogram audio representation by improving upon the baseline. As more temporal structure is added, predictive performance improves. 
    
    This pattern can also be seen in our results with the song split except that predictive performance is better than the frame level baselines at all clip lengths. However, we observe that average performance begins to diminish after ten seconds unlike the album split. This suggests that although there is benefit in the additional temporal data, the model may be overfitting in the song split or that benefits from having a larger training set with many short independent samples are outweighing temporal value. These explanations also imply that excessive temporal information is lost, likely to early pooling layers, when a sample passes through the network. Another interpretation involves the minor translational invariance associated with pooling. While this is desirable in image classification tasks and necessary for computational reasons, it may be detrimental when spatial location has meaning relevant to the classification. 
    
    The discrepancy between song and album splits also verifies the producer effect: test performance is much better when evaluating on unheard songs versus unheard albums. It is important to note that the way an album is produced may also be considered part of an artist's style and is meaningful in certain contexts. This is especially relevant for industry applications such as copyright protection where the producer effect should be taken advantage of as an additional source of discriminative information. Therefore, for a general-purpose framework, strong performance with both types of split is desirable.

    \begin{table}
    \caption{Test F1 Scores for frame level Audio Features (3 runs)} 
    \centering
    \begin{tabular}{|c|c|c c c c c c|}
    \hline
    \textbf{Split} & \textbf{Type} & \textbf{1s} & \textbf{3s} & \textbf{5s} & \textbf{10s} & \textbf{20s} & \textbf{30s }\\
    \hline
    \hline
    Song & Average & 0.729 & 0.765 & 0.770 & \textbf{0.787} & 0.768 & 0.764 \\
    Song & Best & 0.733 & 0.768 & 0.779 & 0.772 & \textbf{0.792} & 0.771 \\
    \hline
    Album & Average & 0.482 & 0.513 & 0.536 & 0.538 & 0.534 & \textbf{0.603} \\
    Album & Best & 0.516  & 0.527 & 0.550 & 0.560 & 0.553 & \textbf{0.612} \\
    \hline
    \end{tabular}
    \label{results_frame}
    \end{table}

    \begin{figure*}
        \centering
        \includegraphics[width=1\textwidth]{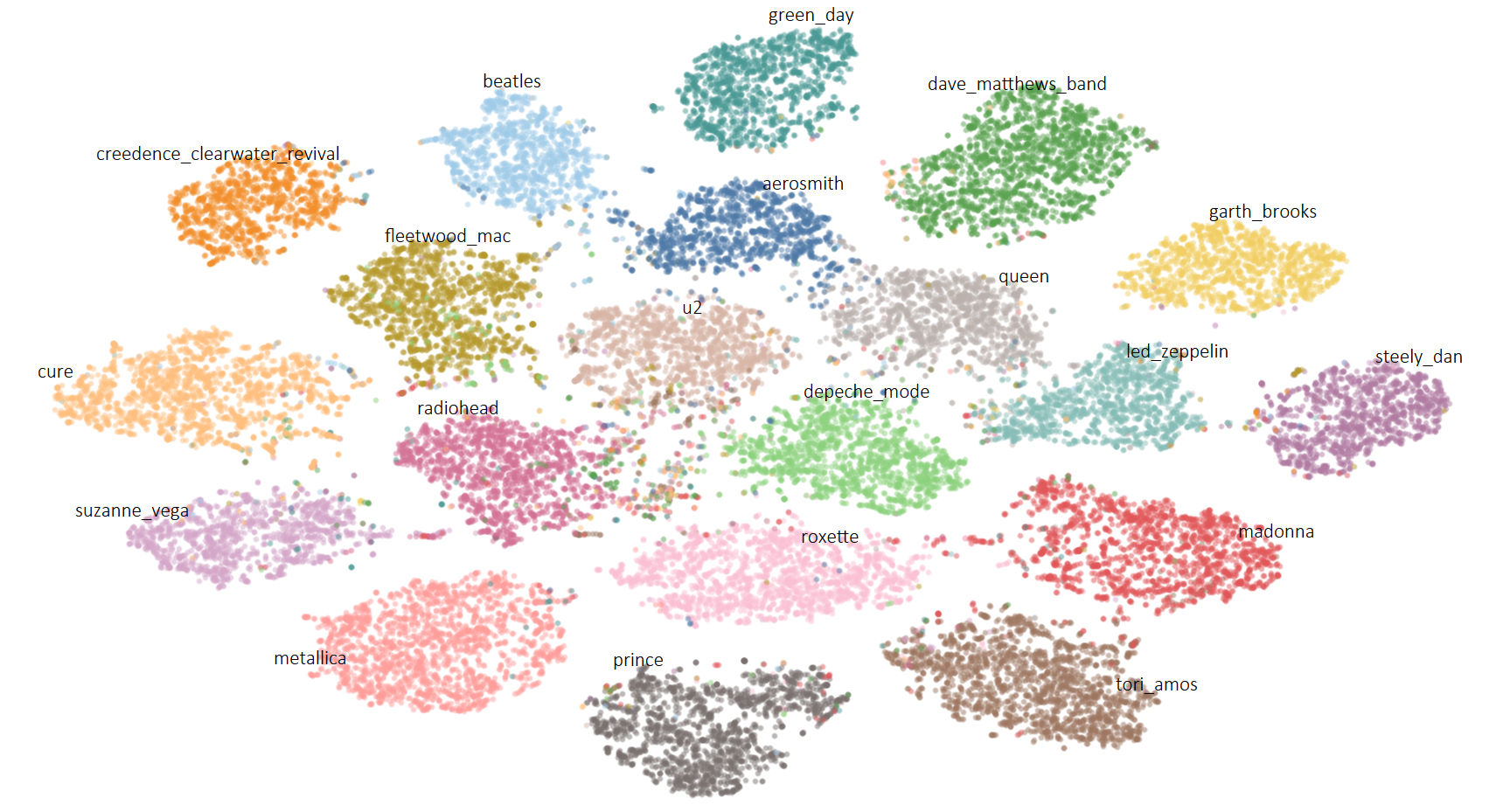}
        \caption{\label{fig:tsne}t-SNE of learned audio representations using audio length of ten seconds (frame level; song split)}
        \label{audio}
    \end{figure*}
        
    \subsection{Song Level Evaluation}
    
    Among baselines, the best song level classification performance is obtained by Mandel and Ellis \cite{mandel2005song} of 0.687 and 0.839 for album and song splits respectively. This is an improvement over the frame level baselines and we observe similar results (Table \ref{results_song}) when aggregating frames at the song level.  

    \begin{table}
    \caption{Test F1 Scores for song level Audio Features (3 runs)} 
    \centering
    \begin{tabular}{|c|c|c c c c c c|}
    \hline
    \textbf{Split} & \textbf{Type} & \textbf{1s} & \textbf{3s} & \textbf{5s} & \textbf{10s} & \textbf{20s} & \textbf{30s }\\
    \hline
    \hline
    Song & Average & 0.929 & \textbf{0.937} & 0.918 & 0.902 & 0.861 & 0.846 \\
    Song & Best & 0.944 & \textbf{0.966} & 0.930 & 0.915 & 0.880 & 0.851 \\
    \hline
    Album & Average & 0.641 & 0.651 & 0.652 & 0.630 & 0.568 & \textbf{0.674} \\
    Album & Best & \textbf{0.700}  & 0.653 & 0.662 & 0.683 & 0.609 & 0.691 \\
    \hline
    \end{tabular}
    \label{results_song}
    \end{table}
   
    In terms of an album split, our results are comparable to Mandel and Ellis at most audio lengths. Song level evaluation improves predictive performance in all cases and this may be attributed to the variance reduction effect of voting. There also does not appear to be any relationship between audio length and predictive performance suggesting that additional temporal structure is of limited value. 
    
    With a song split, the results are more pronounced. Although all audio lengths see a performance gain and outperform the baseline, shorter audio clips observe a much larger boost in comparison. The best-performing three second (3s) case achieves average and best test F1 scores of 0.937 and 0.966 respectively. This is an absolute increase of 10\% over the most comparable baseline (Mandel) and such a model could be immediately useful for real-world applications. As the audio length is increased beyond three seconds though, performance begins to diminish and this is likely because the noise rejection from voting using a larger number of test samples outweighs additional temporal benefit. However, using too short of an audio clip limits the model's capacity to discern artists while longer clips reduce the potential to mitigate wrong predictions through voting. In practice, we advise using an artist's discography for training and song level evaluation with three to ten second long audio samples when possible. 
    
    Overall, from the four conditions resulting from split type and feature level, the CRNN model outperforms the most comparable baseline for at least one audio clip length. This holds true for both best and average case scores except for the album split with song level features where the CRNN model only outperforms in its best-run. This discrepancy may be explained by Mandel's dataset containing fewer classes or by considering that we, unlike the baselines works, report the average of three independent trials instead of performance on a single trial. 
    
    \subsection{Audio Representation Visualization}
    
    At the bottleneck layer of the network, the layer directly proceeding final fully-connected layer, each audio sample has been transformed into a vector which is used for classification. These vectors describe representations learned by the model to distinguish classes. Using t-Distributed Stochastic Neighbor Embedding (t-SNE) \cite{maaten2008visualizing}, one can further reduce dimensionality to visually explore class separation. This is shown in Figure \ref{fig:tsne} for the model trained with audio length of ten seconds. Using the ground truth to color audio samples, the figure demonstrates that the convolution and recurrent layers in the CRNN model create an effective data representation for discriminating artists. Even in two dimensions, most audio samples are well separated and form clusters which uniquely describe a specific artist. While noisy samples exist, this may be attributed to the fact that artists will have songs or at least audio segments within songs which are similar to one another. Furthermore, although the axes are not directly interpretable, some high-level patterns do exist. For example, samples from rock bands such as the Beatles, Aerosmith, Queen and Led Zeppelin project into a similar neighborhood whereas individual pop artists, such as Madonna and Tori Amos, project into a different neighborhood.

    \subsection{Limitations and Future Directions} \label{limits}
    A key assumption in artist classification is that each artist has a unique style which can be learned by a model. If false, it is not possible to generalize well. This can happen in any of the following scenarios: an artist's style is heavily influenced by another, drastic changes in musical style or collaborations where artists feature on each others' tracks. These issues become increasingly complex with electronic or pop music where songs may be stylistically similar and vocals are frequently provided by featuring artists. However, an artist classification model can still be useful despite these conditions if it is primarily being used to measure similarity. An interesting future direction involves extending the visualization portion of this work to verify whether the learned representations of songs organically form clusters in their feature space based on genre and stylistic similarity.

    Another important consideration for artist classification is that there are a limited number of songs belonging to each artist. A way around this is to augment the original audio. For example, Salamon \cite{salamon2017deep} shows that audio augmentation strategies (time stretching, pitch shifting and dynamic range compression) improve accuracy in audio classification tasks. This would also be beneficial for ensuring the model is robust to audio manipulation in industry applications. One could also modify the spectrogram slicing method used in this work to include overlapping windows, rather than disjoint ones, which would expose the model to more examples of temporal structure. Another approach to explore is pre-training the model for another objective, such as genre classification where ample training samples are available, before fine-tuning to perform artist classification. This allows early layers in the network to learn low-level audio features \cite{choi2016explaining} which encourages the model to focus on high-level structure during fine-tuning. Implementing any of these strategies would be a promising direction for future research. 
    
    As mentioned in the results, it is hypothesized that early pooling layers may be discarding too much temporal information. A good direction to extend this work would be to experiment with more temporal data entering the recurrent portion of the network. 
    
\section{Conclusion}
This paper establishes a deep learning baseline for music artist classification on the \textbf{\textit{artist20}} dataset and demonstrates that a Convolutional Recurrent Neural Network is able to outperform traditional baselines under a range of conditions. The results show that including additional temporal structure in an audio sample improves classification performance and also that there is a point beyond which the returns may diminish. This is attributed to a possible lack of complexity in the model or early pooling layers discarding too much information. Using the trained models, predictions are also aggregated at the song level using a majority vote to determine the artist performing a song. This leads to another substantial gain in performance and validates the feasibility of using a CRNN for industry applications such as copyright detection. The best-performing model is trained using three second audio samples under a song dataset split and evaluated at the song level to achieve an average F1 score of 0.937 across three independent trials. Additionally, we visualize audio samples at the bottleneck layer of the network to show that learned representations cluster by artist---highlighting the model's capability as a feature extractor. Future directions include audio augmentation, model pre-training and minimizing temporal pooling as avenues for further performance improvement.

\vspace{12pt}
\bibliographystyle{IEEEtran}
\bibliography{ref}

\end{document}